\documentclass{cimento}
\usepackage{multirow,mathtools,amsmath,amssymb,graphicx,epsfig,amsfonts,color,epstopdf,adjustbox,lipsum,slashed,float,subfloat}

\title{Right-Handed Neutrinos: DM and LFV $vs$.
Collider}
\author{M.~Chekkal\from{ins:x},
A.~Ahriche\from{ins:x2},
A.B.~Hammou\from{ins:x}
 \atque
S.~Nasri\from{ins:x3}}

\instlist{\inst{ins:x} Department of Physics, University of Sciences and Technology of Oran,
BP 1505, Oran, El M'Naouer, Algeria.
 \inst{ins:x2} Department of Physics, University of Jijel, PB 98 Ouled Aissa, DZ-18000
Jijel, Algeria.
 \inst{ins:x3} Department of physics, United Arab Emirates University, Al-Ain, UAE.}

\begin{document}

\maketitle

\begin{abstract}
In a class of neutrino mass models with a lepton flavor violation (LFV) Yukawa interaction term that involves a heavy right handed neutrino, a charged scalar and a charged lepton, we investigate at the ILC@500 GeV the possibility of observing news physics. These models can address neutrino mass and dark matter without being in conflict with different LFV constraints. By imposing DM relic density and LFV constraints, we recast the analysis done by L3 collaboration at LEP-II of monophoton searches on our space parameter and look for new physics in such channels like monophoton and $S S(\gamma)$, where we give different cuts and show the predicted distributions. We show also that using polarized beams could improve the statistical significance.
\end{abstract}

\section{Introduction}

Neutrino oscillations have been put in evidence by different experiences and observations caused by nonzero neutrino masses and neutrino mixing~\cite{exp}. However, the Standard Model (SM) does not explain the intrinsic properties of neutrinos such as their origin, nature and the smallness of their masses. The seesaw mechanism~\cite{seesaw} is the most popular method to explain the tiny mass of SM neutrinos but poses scale problems and prevents the direct detection of the right-handed (RH) neutrinos introduced by this mechanism because of its large mass compared to the electroweak scale. \\
The radiative neutrino mass models~\cite{Zee,ZB,Ma,KNT,Aoki:2008av} is another way to generate a small mass to light neutrinos at loop level and to circumvent the scale problem. The violation of the leptonic number is permitted by the fact that the neutrinos are Majorana particles where the lightest RH neutrino is identified as being the dark matter and have large phenomenological implications. We can take as an example the model in~\cite {AMN} where authors show that the scale of new physics can be in the sub-TeV for the 3-loops neutrino mass generation model~\cite{KNT} which makes it testable at collider experiments~\cite{ANS}. In this work, we study the possibility of detecting the manifestations of the new physics resulting from this class of radiative neutrino mass models.

\section{LFV and DM Constraints Class of Models with RH Neutrinos}

A class of radiative neutrino mass models are considered here, wich are extending
the SM with three right-handed neutrinos $N_{i}~(i=1,2,3)$
and a $SU(2)_{L}$-singlet charged scalar $S^{\pm}$.
The models contain the following Yukawa term in in the Lagrangian~\cite{Ma,KNT,Aoki:2008av,Okada:2015hia}
\begin{equation}
\mathcal{L_{N}}\supset-\frac{1}{2}m_{N_{i}}\overline{N_{i}^{c}}P_{R}N_{i}+g_{i\alpha}S^{+}\overline{N_{i}}\ell_{\alpha_{R}}+\mathrm{h.c.},\label{LL}
\end{equation}
where $\ell_{\alpha_{R}}$ is the right-handed charged lepton and $g_{i\alpha}$ are Yukawa couplings. The stability of the 
lightest RH neutrino, wich is supposed play the DM role, is assured by imposing the global $Z_{2}$ symmetry\footnote{The 
global $Z_2$ symmetry is accidental for higher representation (setplet) of RH neutrinos~\cite{Ahriche:2015wha}.}. $\ell_{\alpha}\rightarrow \ell_{\beta}\gamma$ 
and $\ell_{\alpha}\rightarrow 3\ell_{\beta}$ are LFV processes produced by this type of interaction.

The contribution of the interactions (\ref{LL}) to the $\ell_{\alpha}\rightarrow\ell_{\beta}\gamma$ branching ratio is given by~\cite{Toma:2013zsa}

\begin{equation}
\mathcal{B}^{(N)}(\ell_{\alpha}\rightarrow\ell_{\beta}\gamma)=\frac{3(4\pi)^{3}\alpha}{4G_{F}^{2}}|A_{D}|^{2}\times\mathcal{B}(\ell_{\alpha}\rightarrow\ell_{\beta}\nu_{\alpha}\bar{\nu}_{\beta}),\label{meg}
\end{equation}
where $\alpha$ is the
electromagnetic fine structure constant, $G_{F}$ is the Fermi constant and $A_{D}$ is the dipole
contribution given by

\begin{equation}
A_{D}=\sum_{i=1}^{3}\frac{g_{i\beta}^{\ast}g_{i\alpha}}{2(4\pi)^{2}}\frac{1}{m_{S}^{2}}F\left(x_{i}\right),\label{AD}
\end{equation}
with $x_{i}=m_{N_{i}}^{2}/m_{S}^{2}$ and $F(x)$ is a loop function.

We scanned over all the free parameters of our model to determine the the phenomenological
implications for the dark matter and the searches of new Physics at colliders. To get a
feeling of the different contributions from the RH neutrino couplings, we define the ratio (fine-tuning parameter)
\begin{equation}
\mathrm{R}=\frac{\mid\sum_{i=1}^{3}g_{i\mu}^{\ast}g_{ie}F\left(x_{i}\right)\mid^{2}}{Max[\mid g_{i\mu}^{\ast}g_{ie}F\left(x_{i}\right)\mid^{2}]},\label{R}
\end{equation}

which represents the way the cancellation between different combinations $g^*_{i\beta}g_{j\alpha}$ occurs
in order to suppress the LFV branching ratios even for large $g$-couplings. For instance, the parameter $R$ could be signiﬁcantly smaller than unity due to possible cancellation between different RH neutrinos contributions, and this may allow the $g$-couplings to be relatively large.

In Fig.~\ref{gg}, for different values of fine-tuning parameter $R\approx ~1,~10^{-2},~10^{-4}$, the branching ratios for 
the processes $\ell_{\alpha}\rightarrow\ell_{\beta}\gamma$ and $\ell_{\alpha}\rightarrow 3 \ell_{\beta}$ versus the charged 
scalar mass are presented. The experimental bounds of all branching fractions are the first constraint to be obeyed by the 
free parameters of the model, and some hundreds of configurations of free parameters are generated in this way.

{\setlength\textfloatsep{8pt plus 2pt minus 2pt}
\setlength\intextsep{8pt plus 2pt minus 2pt}

\begin{figure}[h]
\begin{centering}
\includegraphics[width=0.31\textwidth]{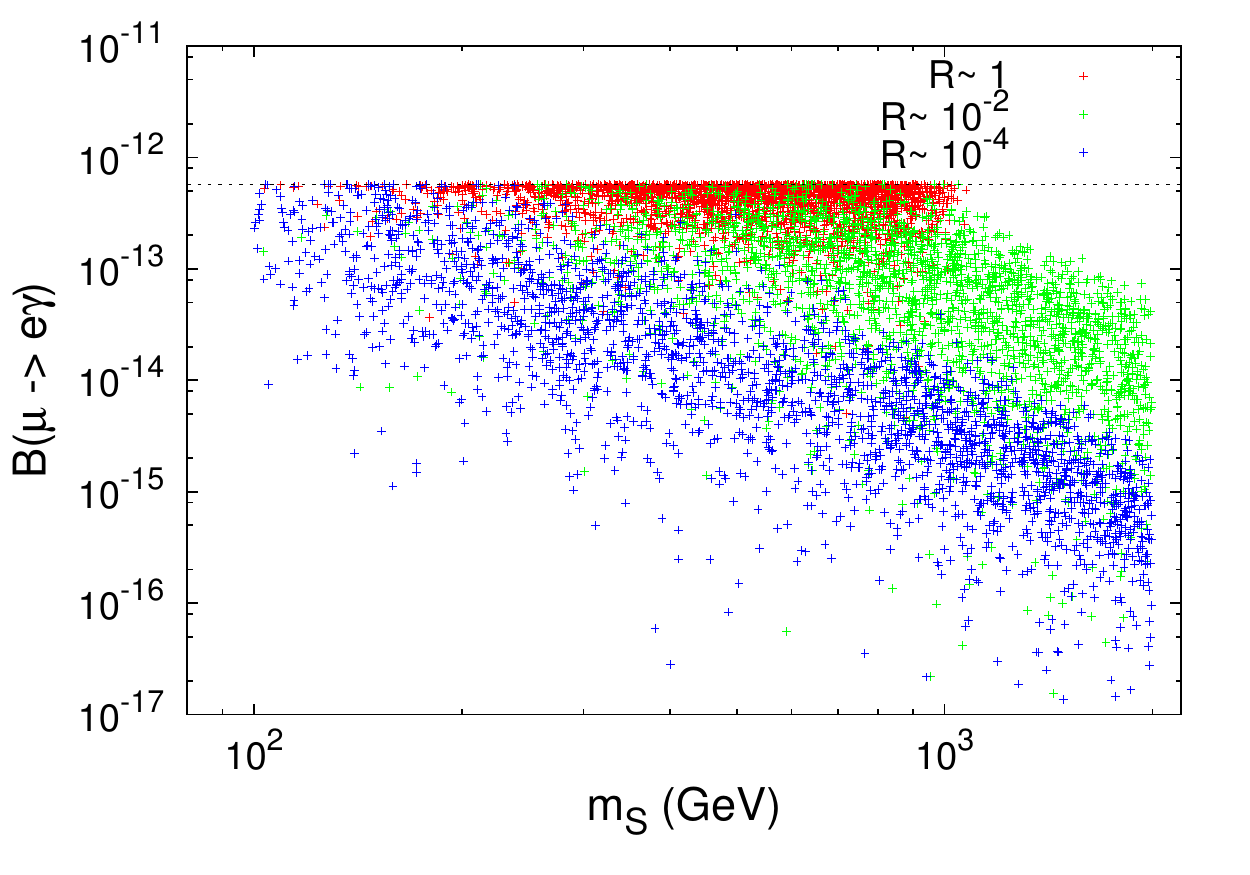}~\includegraphics[width=0.31\textwidth]{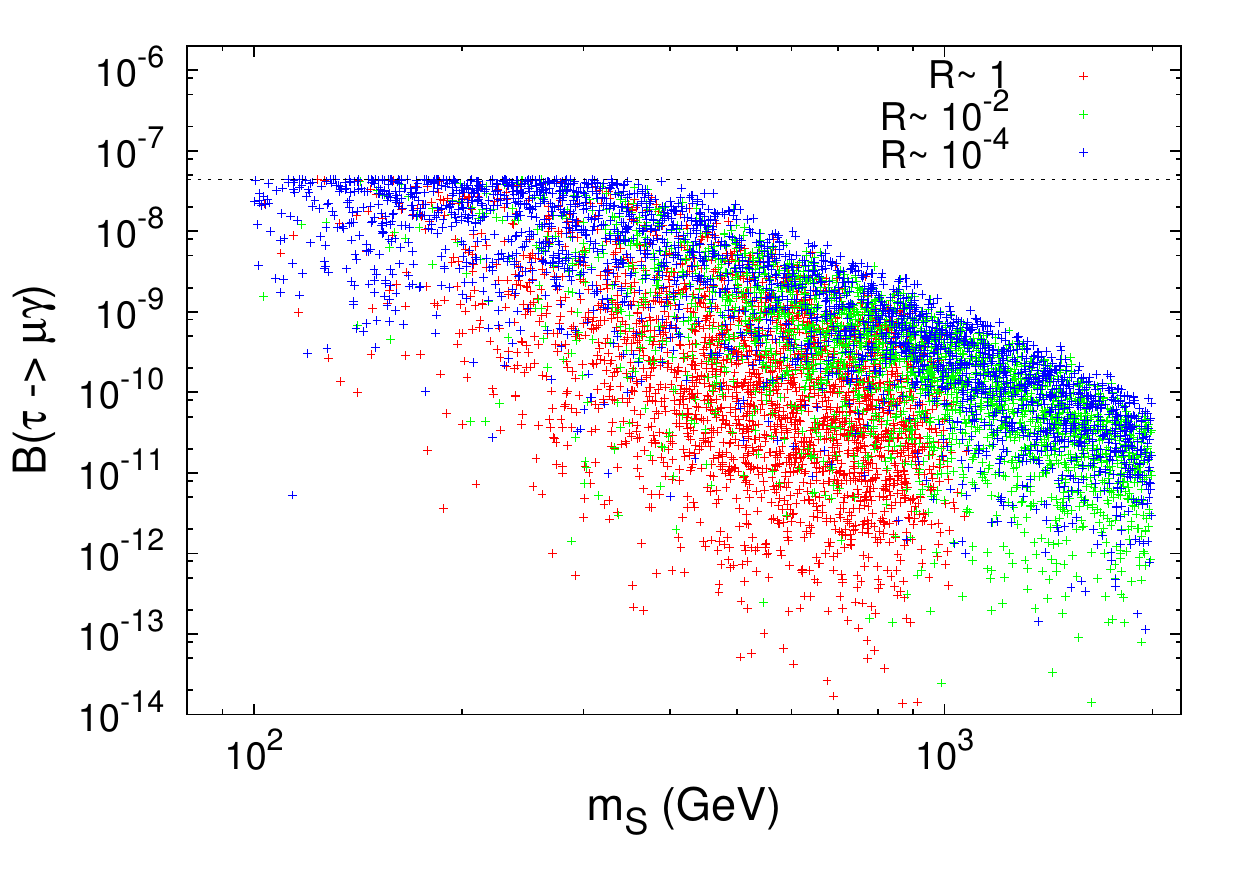}~\includegraphics[width=0.31\textwidth]{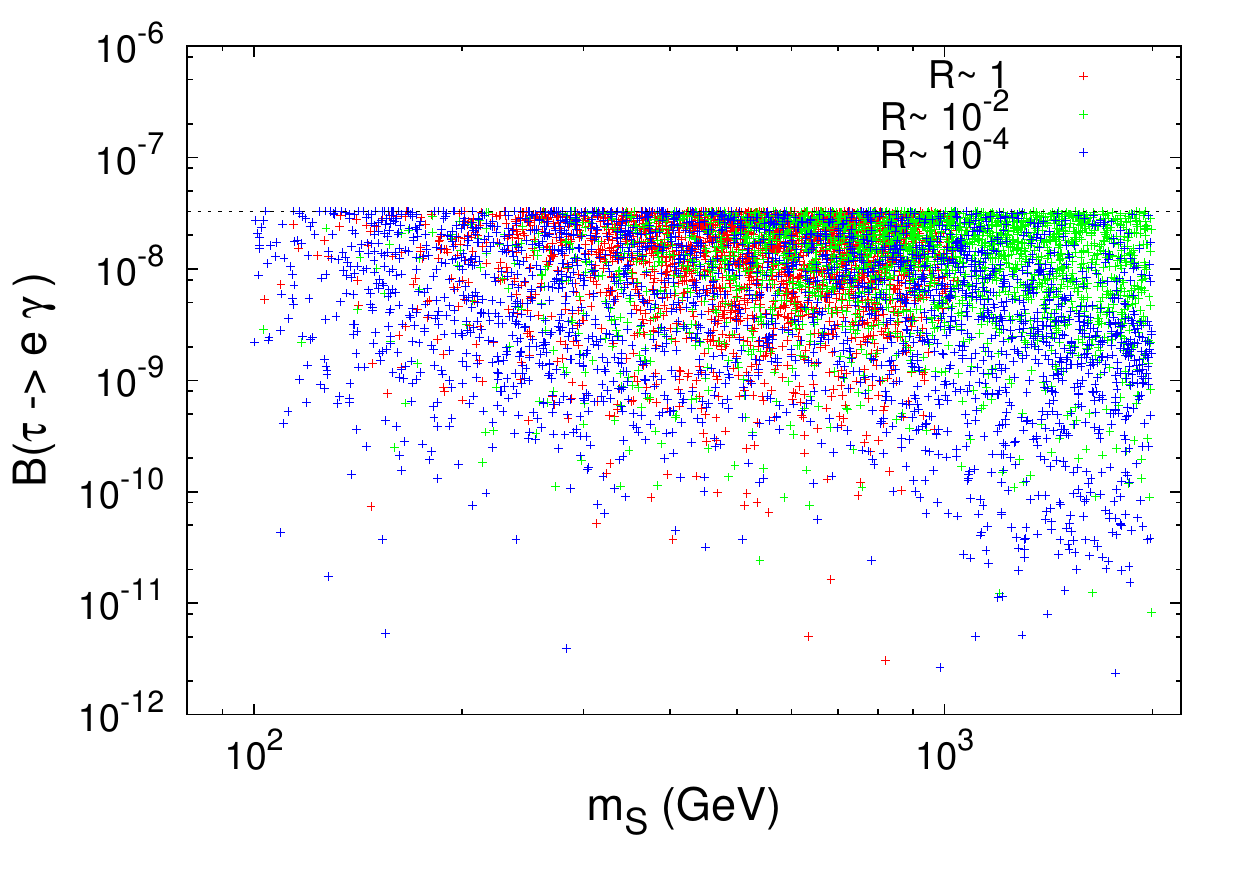}\\
\includegraphics[width=0.31\textwidth]{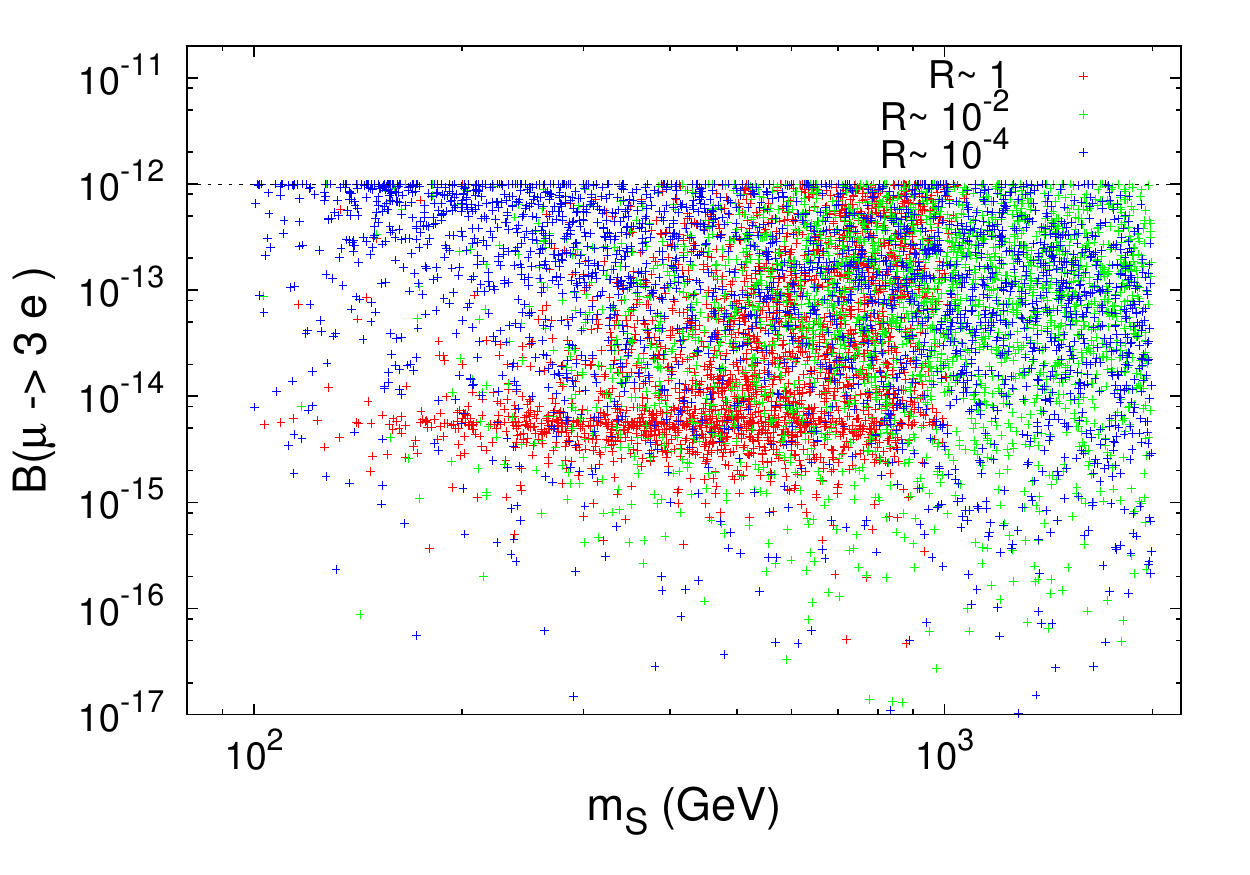}~\includegraphics[width=0.31\textwidth]{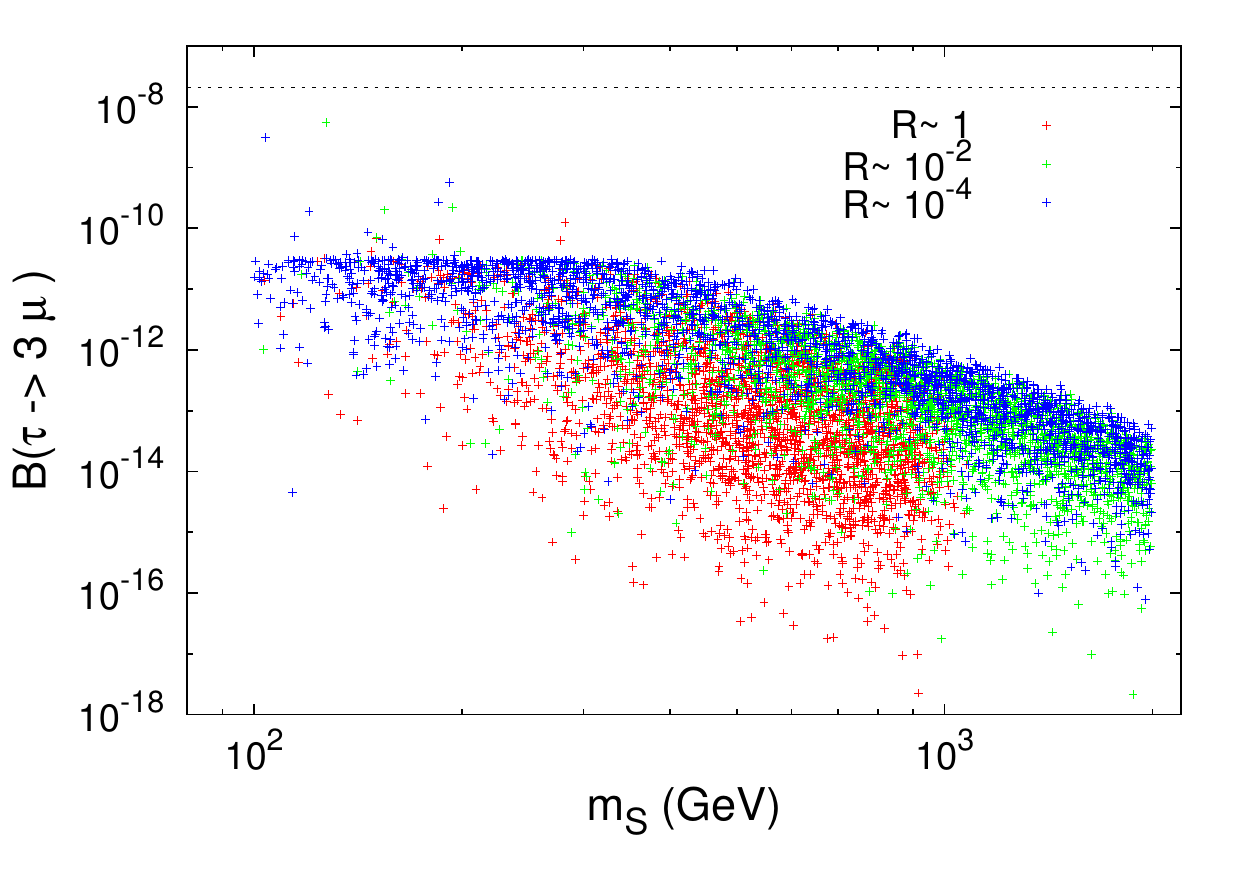}~\includegraphics[width=0.31\textwidth]{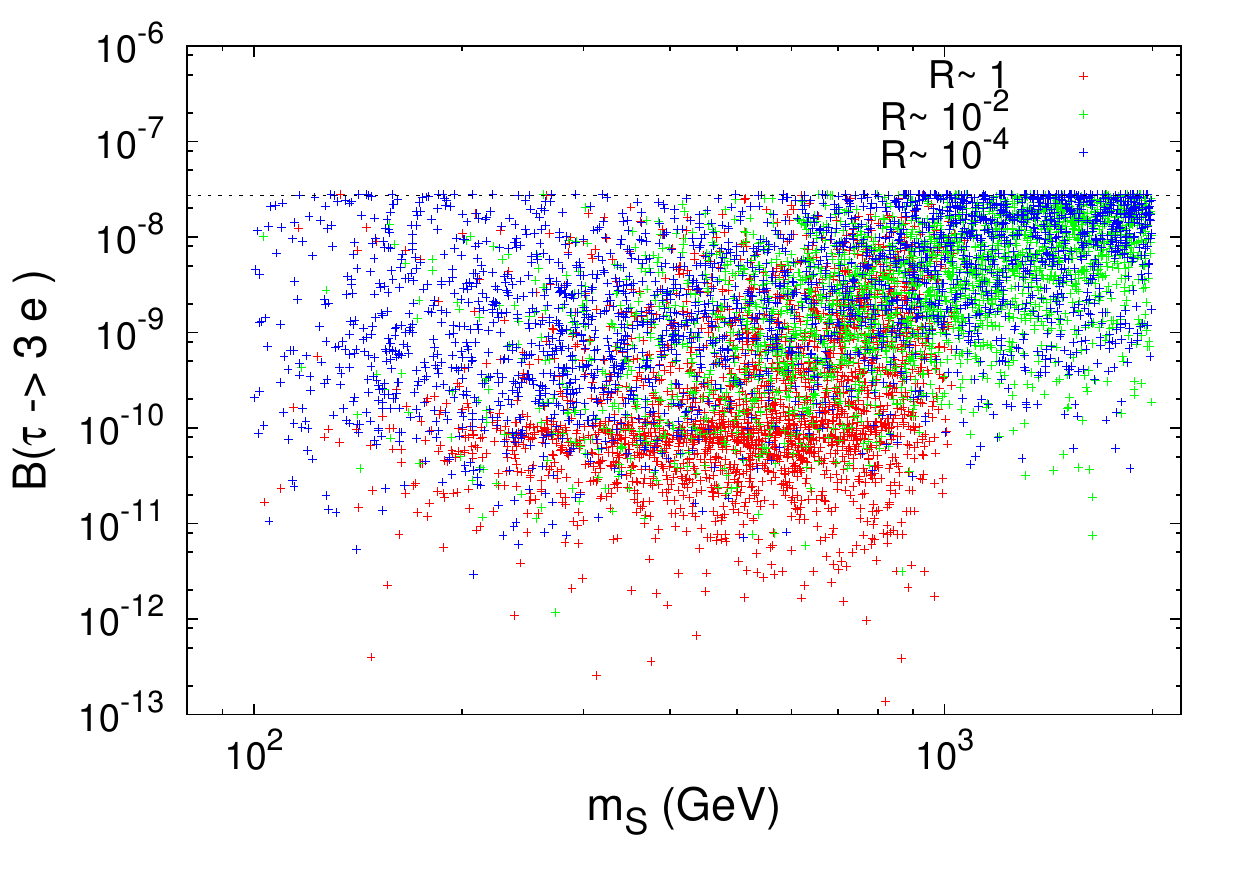}
\par\end{centering}
\caption{The branching ratios (top) $\mathcal{B}(\mu\rightarrow e\gamma)$, $\mathcal{B}(\tau\rightarrow\mu\gamma)$ and $\mathcal{B}(\tau\rightarrow e\gamma)$; and (bottom) $\mathcal{B}(\mu\rightarrow 3e)$, $\mathcal{B}(\tau\rightarrow 3\mu)$ and $\mathcal{B}(\tau\rightarrow 3e)$ versus $m_{S}$. The horizontal dashed lines show the current experimental
upper bounds for each radiative decay.}
\label{gg} 
\end{figure}

As mentioned earlier, the dark matter condidate could be the lightest RH neutrinos $N_1$ which is supposed stable. We can 
safely keep only the contribution of $N_1$ density and neglect that of $ N_2 $ and $ N_3 $ in hierarchical RH neutrino
mass spectrum case. The annihilation process $N_{1}N_{1}\rightarrow\ell_{\alpha}\ell_{\beta}$
via t-channel exchange of $S^{\pm}$ impoverished the density of $ N_1 $. When the temperature of the
universe drops below the freeze-out temperature, and using the equation Boltzmann equation, we can approximate the relic 
density after the decoupling of $ N_1 $ from the thermal bath~\cite{Ahriche:2013zwa}
\vspace{-0.3cm}
\begin{equation}
\Omega_{N_{1}}h^{2}\simeq\frac{2x_{f}\times 1.1\times10^{9}~\text{GeV}^{-1}}{\sqrt{g^{*}}M_{pl}\left\langle \sigma_{N_{1}N_{1}}v_{r}\right\rangle }\simeq\frac{17.56}{\sum_{\alpha,\beta}|g_{1\alpha}g_{1\beta}^{*}|^{2}}\left(\frac{m_{N_{1}}}{50~\text{GeV}}\right)^{2}\frac{\left(1+m_{S}^{2}/m_{N_{1}}^{2}\right)^{4}}{1+m_{S}^{4}/m_{N_{1}}^{4}},\label{rd}
\end{equation}

in Fig.~\ref{r-d}, we present a contour plot $m_{N_{1}}$ versus $m_{S}$, where in palette we have the coupling combination
$\sum_{\alpha\beta}\left|g_{1\alpha}g^{*}_{1\beta}\right|^{2}$, which appears in the expression of the relic density, within 
the conditions $m_{N_{1}}<m_{S}$ and $m_{S}>100$ GeV being imposed. It is difficult to maintain all LFV ratios within the 
current experimental bounds for values of coupling combination larger than 10, and required an extreme fine-tunning. So, once 
the relic density are imposed and $m_{N_1}$ and $m_S$ are defined the $\sum_{\alpha\beta}\left|g_{1\alpha}g_{1\beta}\right|^{2}$ 
impose another condition in addition to LFV constraint and the most viable range of the
masses is extracted as $m_{N_{1}}<200~\text{GeV}$ and $m_{S}<300~\text{GeV}$.

\section{Constraints from LEP-II}

An additional constraint on the free parameters is imposed by the no evidence for massive neutral particle realized by the L3 
detector at LEP-II~\cite{LEP}, which has conducted an analysis on single and multi photon events with missing for center of 
mass energies between 189 and 209 GeV. Indeed, benchmark points that respect the different DM and LFV constraints together 
must also give non-revelent significance under the same conditions as those of LEP.

In the next sections, we will carried out the electron-positron (electron-electron) collision on the ILC, so the decay 
length of the unstable particles $ N_2 $ and $ N_3 $ must then be measured to determine whether they are disintegrating 
inside or outside the detectors. From Fig.~\ref{decl}, It can be seen that $ N_3 $ does not contribute to the missing energy, 
because being disintegrated mainly inside the detectors, whereas a substantial amount of $N_2$ events escape from the detector. 
In all our analysis the benchmarks points are verified in order to precisely identify the missing energy.

\begin{figure}[h]
\begin{centering}
\includegraphics[width=0.48\textwidth,height=0.20\textheight]{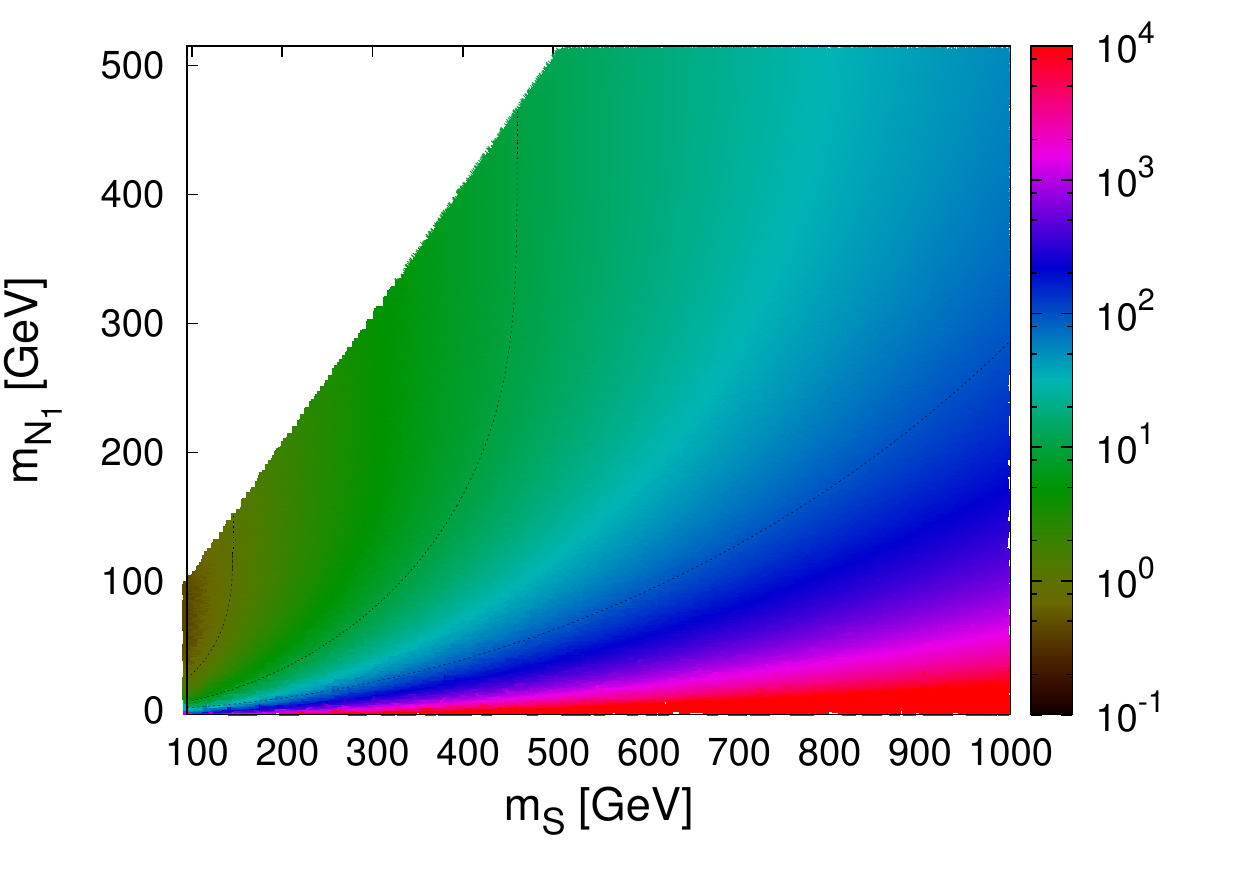} 
\par\end{centering}
\caption{Dark Matter mass versus the charged scalar mass, The palette represents the quantity
$\sum_{\alpha\beta}\left|g_{1\alpha}g^{*}_{1\beta}\right|^{2}$. The dashed curves (from left to right) represent the values 
$\sum_{\alpha\beta}\left|g_{1\alpha}g^{*}_{1\beta}\right|^{2}=1,~10,~100$, respectively.}
\label{r-d} 
\end{figure}

\begin{figure}[htp]
\begin{centering}
\includegraphics[width=0.42\textwidth,height=0.20\textheight]{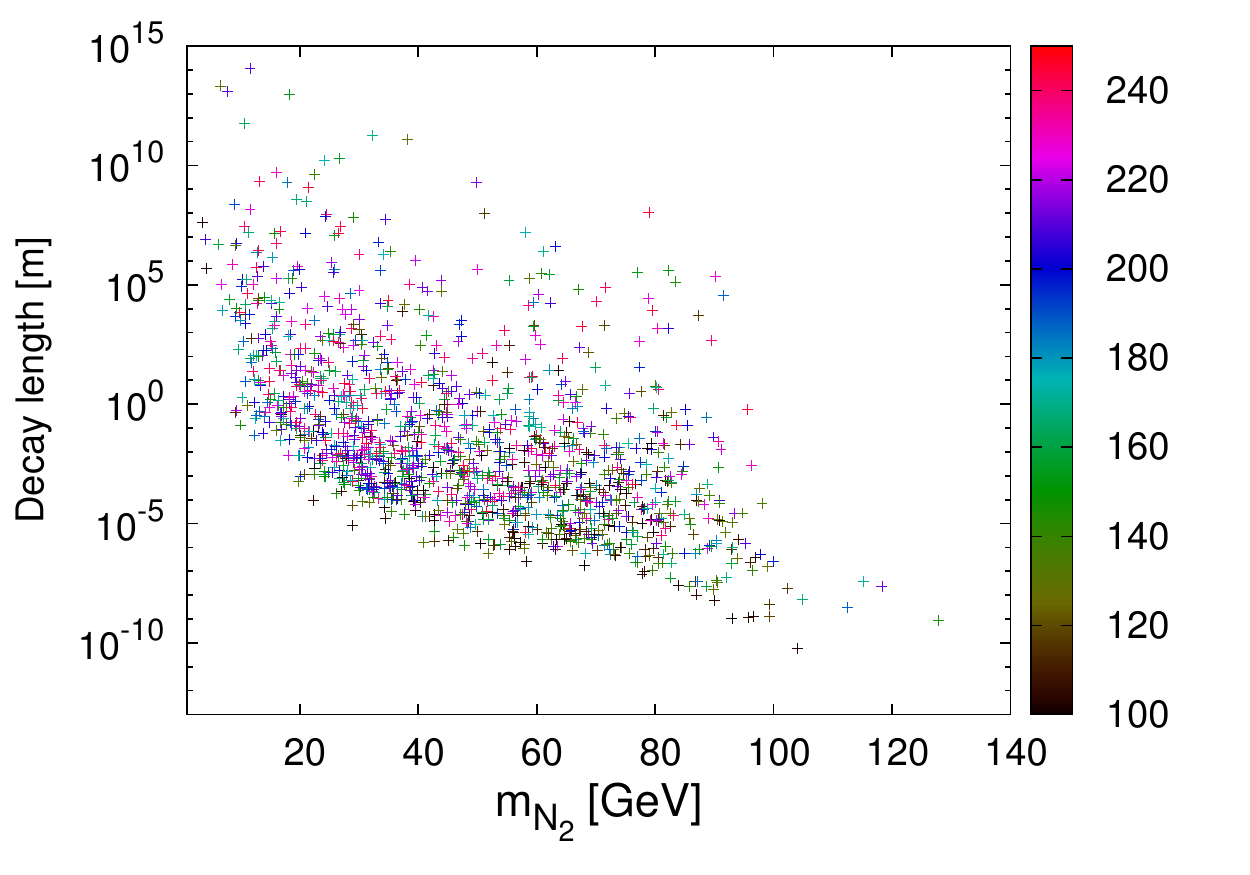}~\includegraphics[width=0.42\textwidth,height=0.20\textheight]{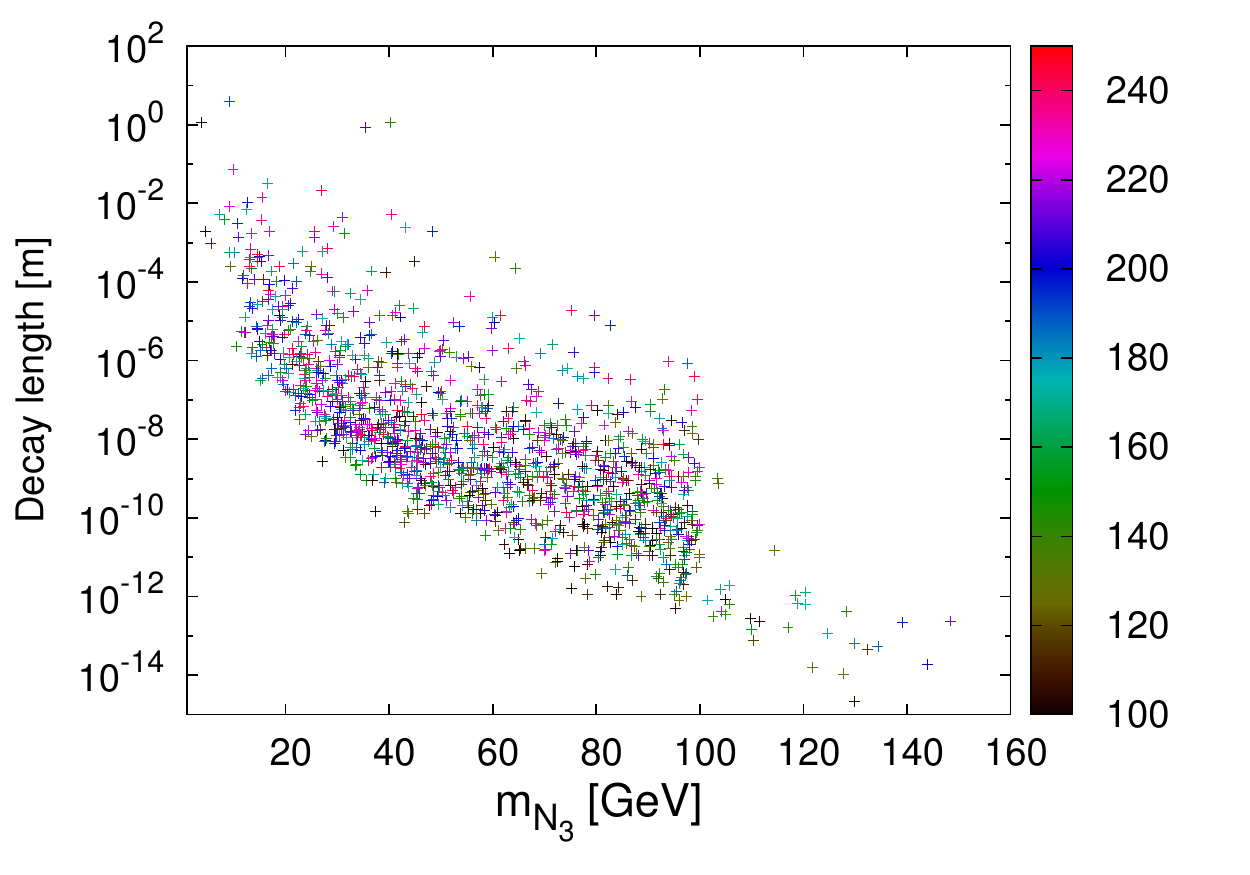} 
\par\end{centering}
\caption{The decay length of the RH neutrinos $N_2$ (left) and $N_3$ (right) as a function of $m_{N_2}$ and $m_{N_3}$, respectively. 
The palette represents the charged scalar mass $m_S$ [GeV].}
\label{decl} 
\end{figure}

We consider the highest integrated luminosities $176~\text{pb}^{-1}$ and $130.2~\text{pb}^{-1}$ at the center of mass energies 
$\sqrt{s}=188.6~\text{GeV}$ and $\sqrt{s}=207~\text{GeV}$, respectively. Same kinematical cuts used by L3 collaboration for a 
high energy single photon are applied~\cite{LEP}: $\left|cos\,\theta_{\gamma}\right|<0.97 $, $p_{t}^{\gamma}>0.02\sqrt{s} $ and 
$E_{\gamma}>1\,\text{GeV}$. We compute the cross sections of the signal $e^{-}e^{+}\rightarrow\gamma+E_{miss}$ and the background 
$e^{-}e^{+}\rightarrow\nu_{i}\bar{\nu}_{j}\gamma$ using the LanHEP/CalcHEP packages~\cite{LanHEP,CalcHEP}, for thousands of 
aforementioned benchmark points.

\begin{figure}[h]
\begin{centering}
\includegraphics[width=0.38\textwidth,height=0.18\textheight]{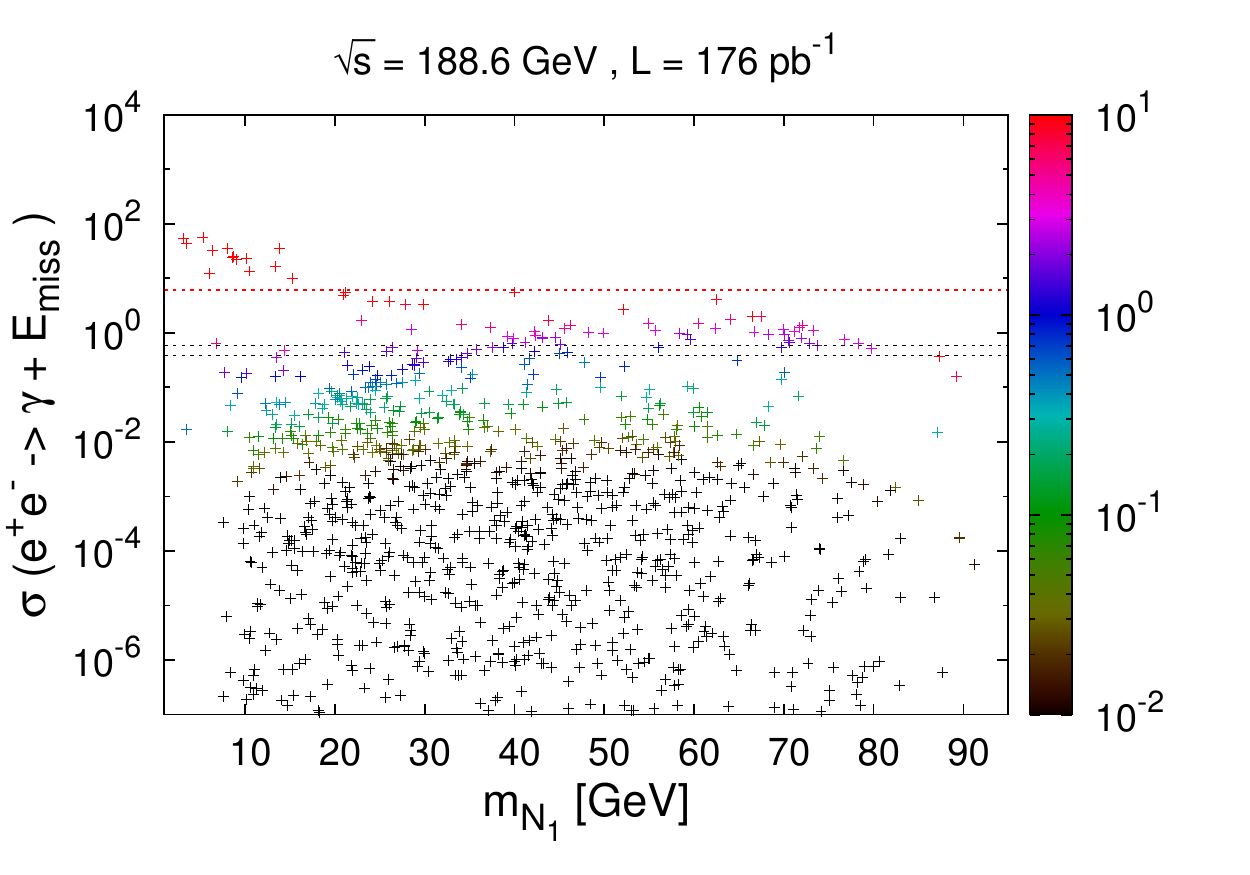}~\includegraphics[width=0.38\textwidth,height=0.18\textheight]{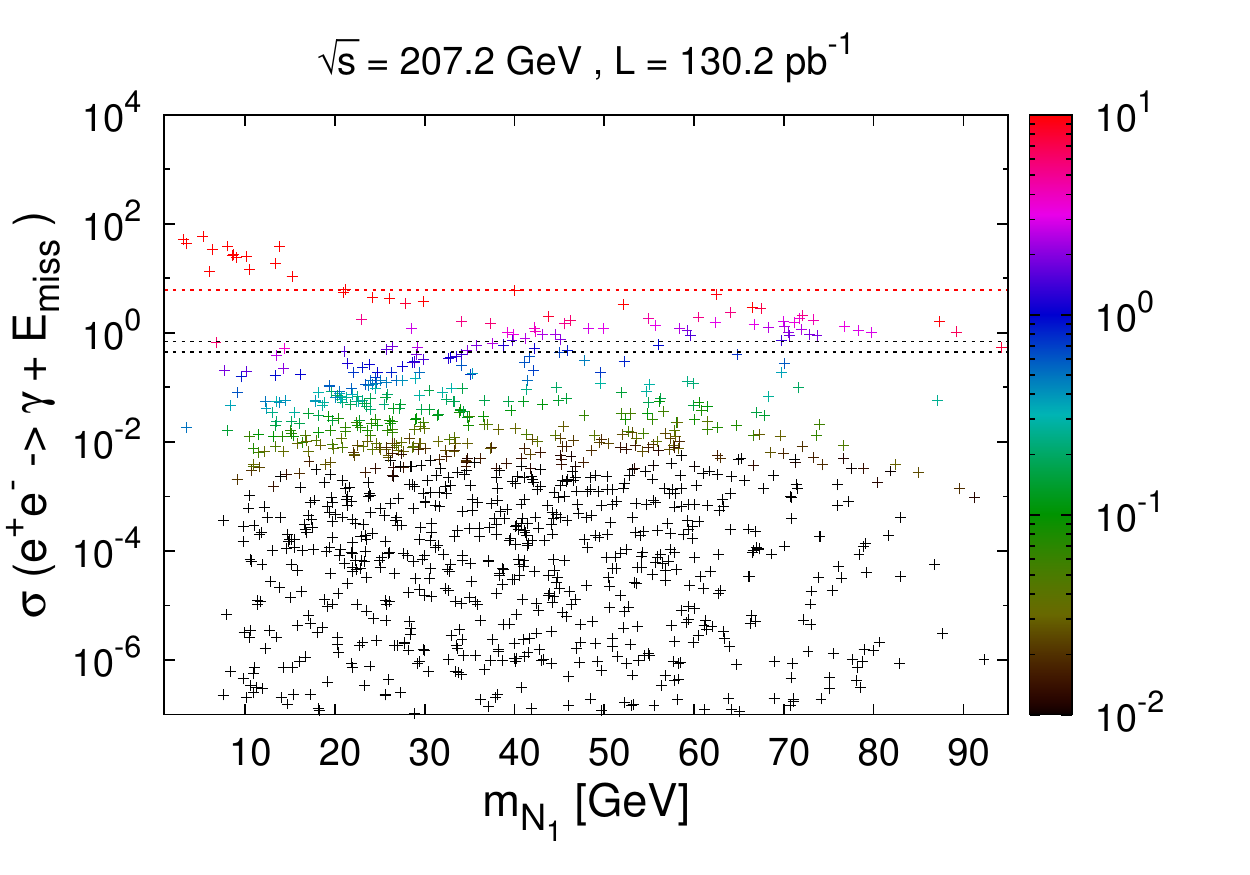} 
\par\end{centering}
\caption{The cross section for the randomly chosen benchmark points for the process $e^{-}e^{+}\rightarrow\gamma+E_{miss}$ 
at LEP as a function of $m_{N_{1}}$ for the CM energies $\sqrt{s}=188.6~\text{GeV}$ (right) and $\sqrt{s}=207.2~\text{GeV}$ 
(left). The palette represents the combination $\Delta$, and the black dashed lines correspond to $S=2,~3$, respectively. The red dashed 
line corresponds to the background.}
\label{lep} 
\end{figure}

The results are shown in Fig.~\ref{lep}, where in palette one can reads $\Delta$, a quantity at which the cross-section is 
sensitive. An exclusion bound on a combination of these parameter is drived according to LEP analysis and the significance $S$ 
must be smaller than three, thereby we extract the following constraint,

\begin{equation}
\Delta =\sum_{i,k} \left|g_{ie}g_{ke}^{*}\right|^{2} \left[\frac{150~\text{GeV}}{m_{S}}\right] \left[\frac{50~\text{GeV}}{\sqrt{m_{N_i}m_{N_k}}}\right] <1.95.\label{LEP}
\end{equation}

\section{Possible Signatures at Lepton Colliders}

In this work, we are interested in the possibility of probing/detecting the traces of the new physics mediated by the charged 
scalar and giving dark mater in the final state, at lepton collider specially the International Linear Collider (ILC)~\cite{ilc} 
which covers center of mass (CM) energies from $250$ to $500~\text{GeV}$. We study the most interesting signatures 

\begin{figure}[htp]
\begin{centering}
\includegraphics[width=0.32\textwidth]{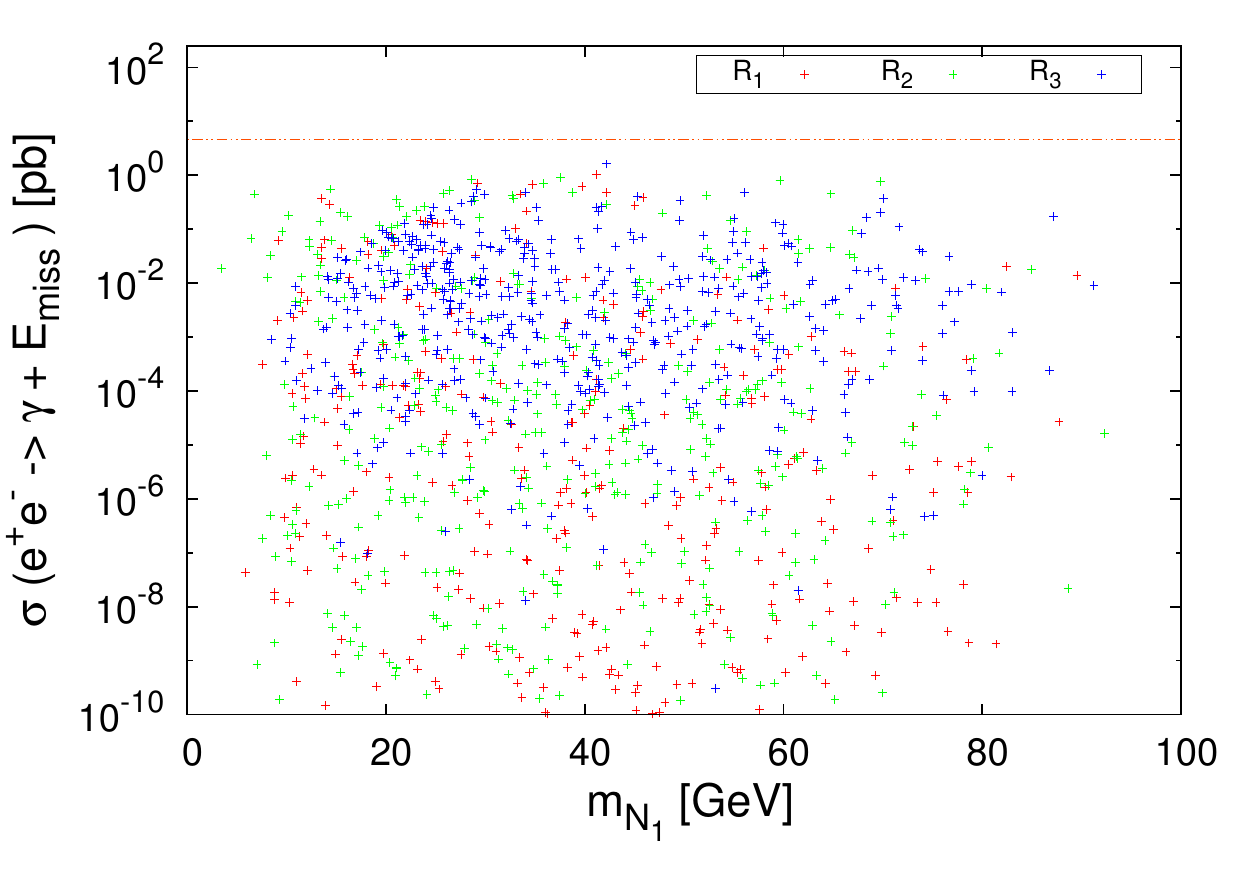}~\includegraphics[width=0.32\textwidth]{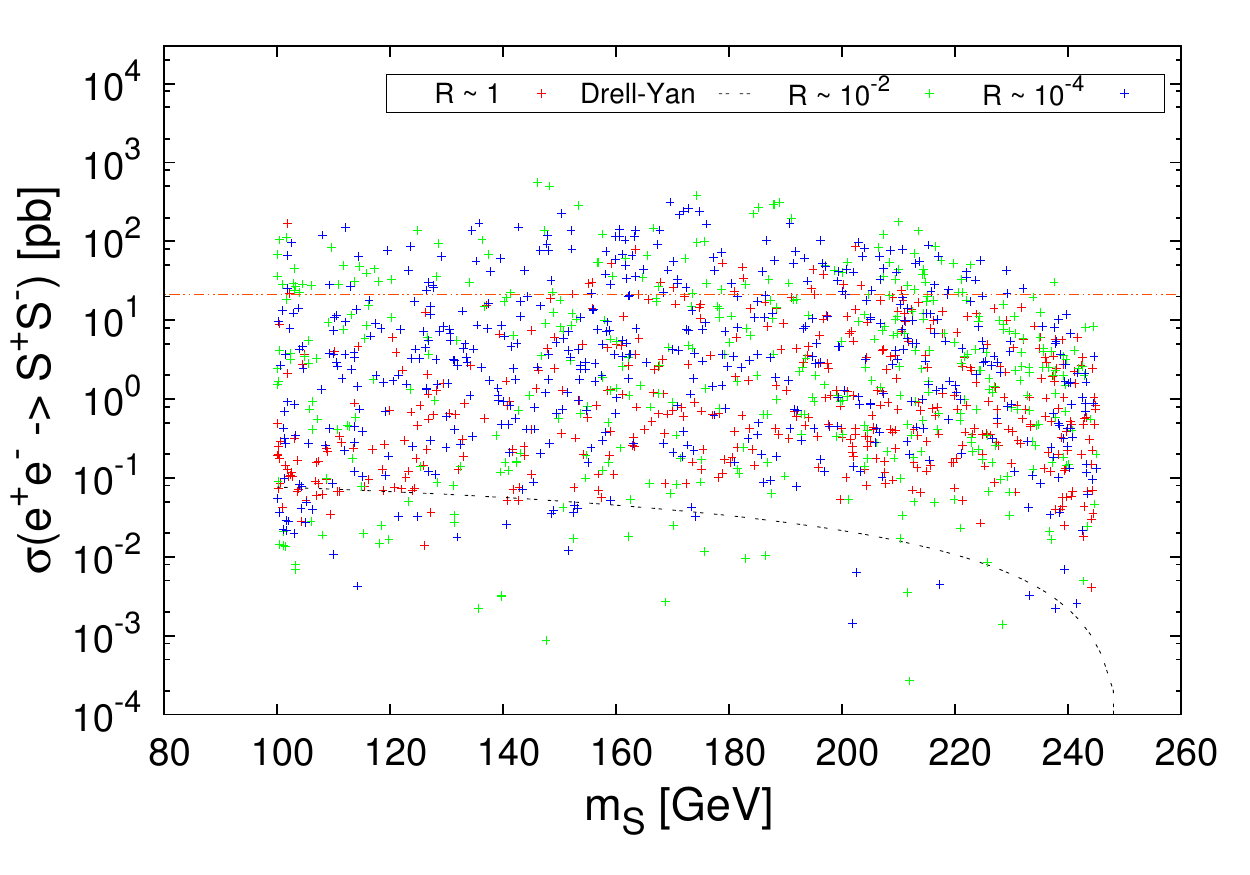}~\includegraphics[width=0.32\textwidth]{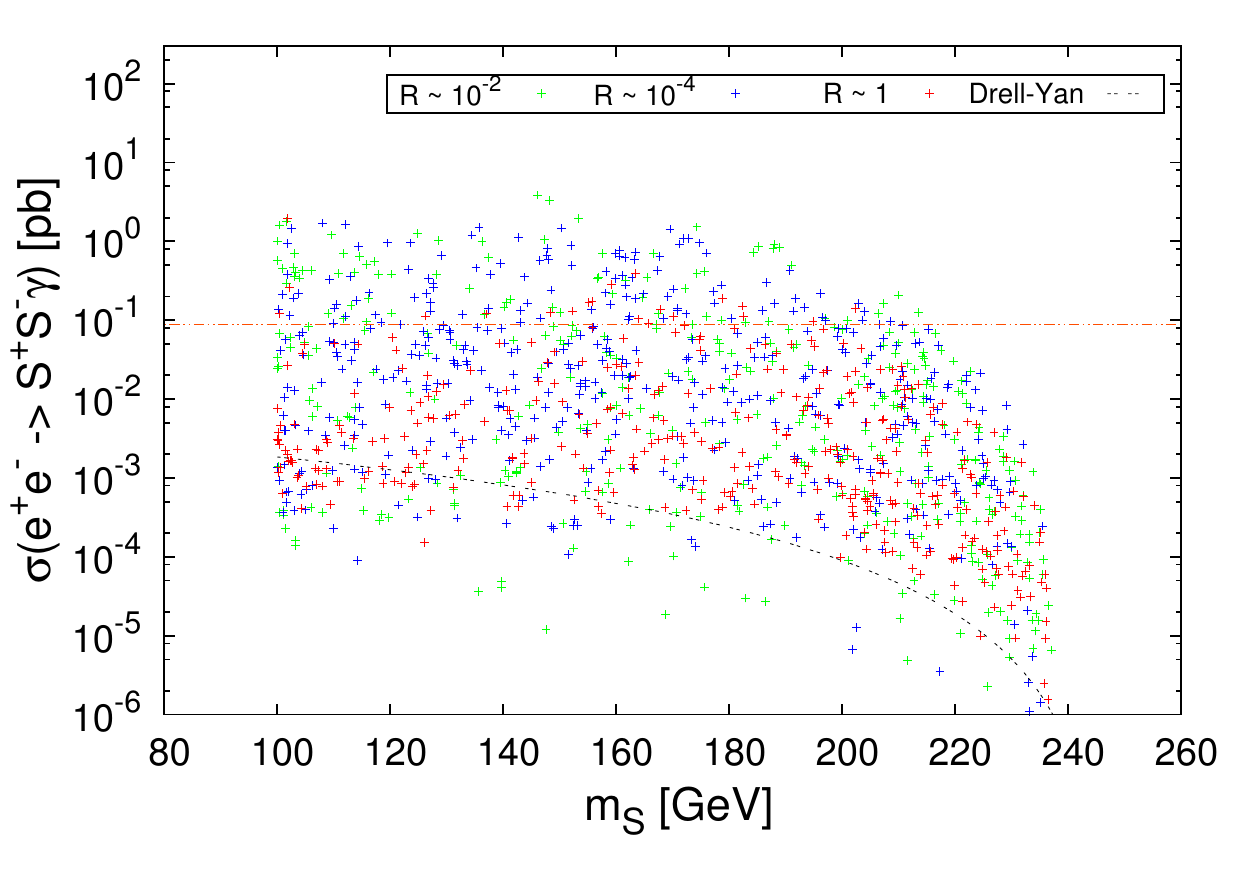}\\
\includegraphics[width=0.32\textwidth]{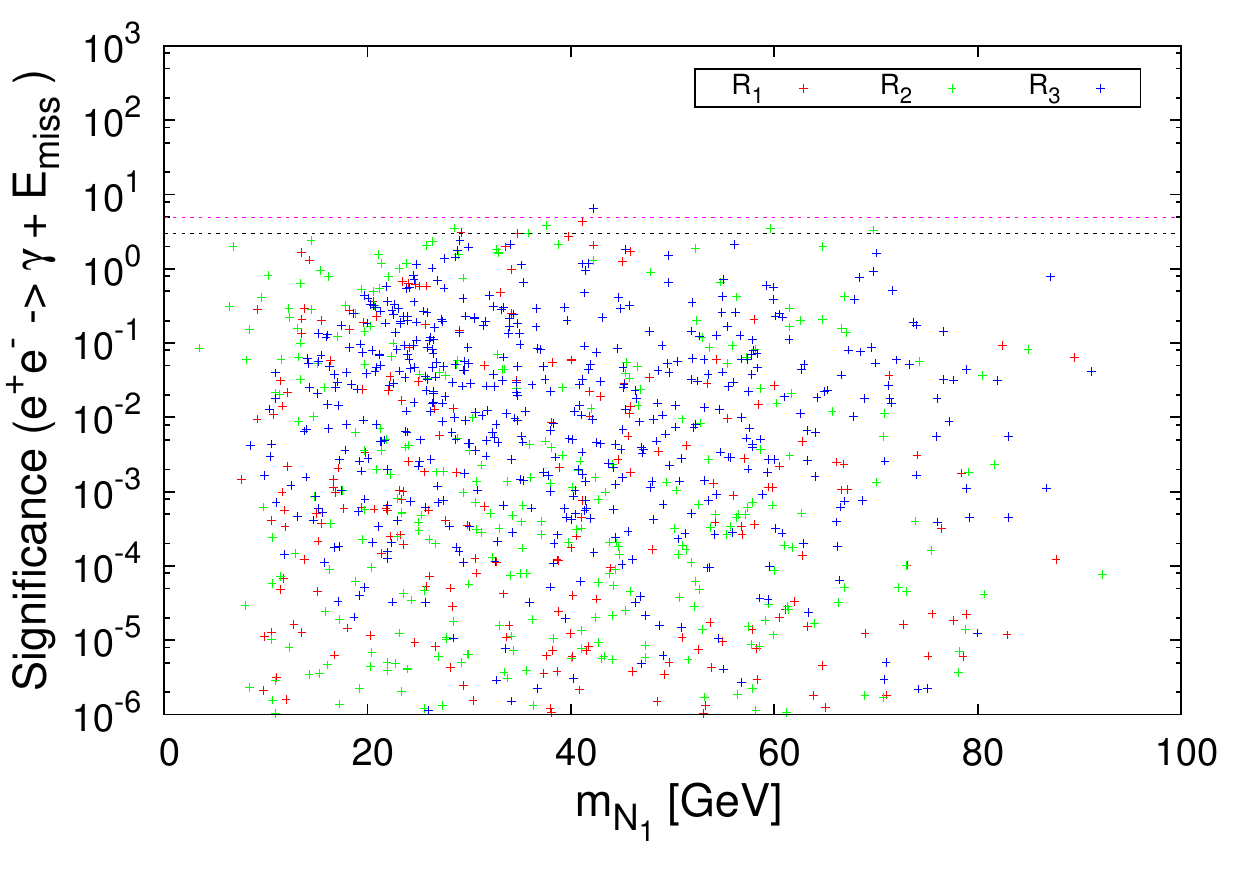}~\includegraphics[width=0.32\textwidth]{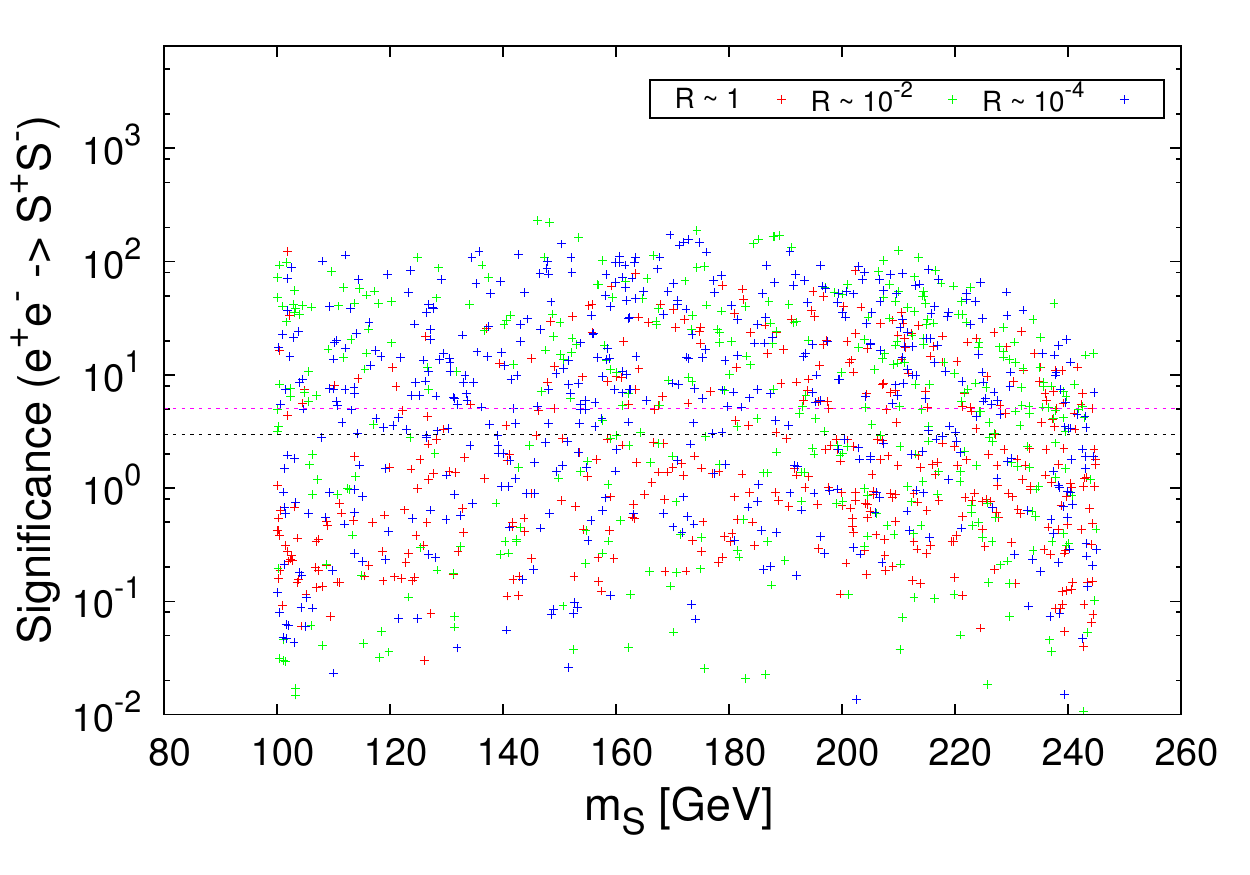}~\includegraphics[width=0.32\textwidth]{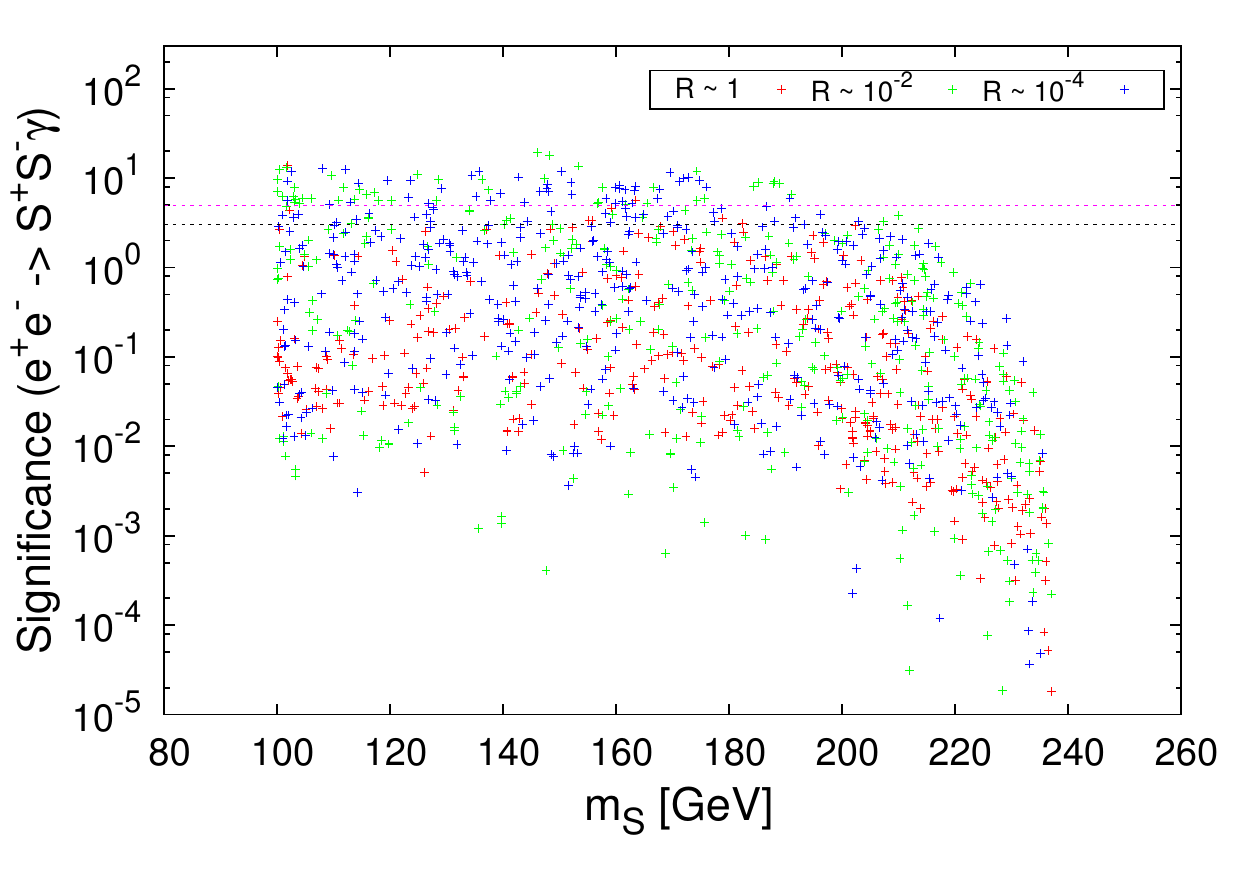}\\
\includegraphics[width=0.32\textwidth]{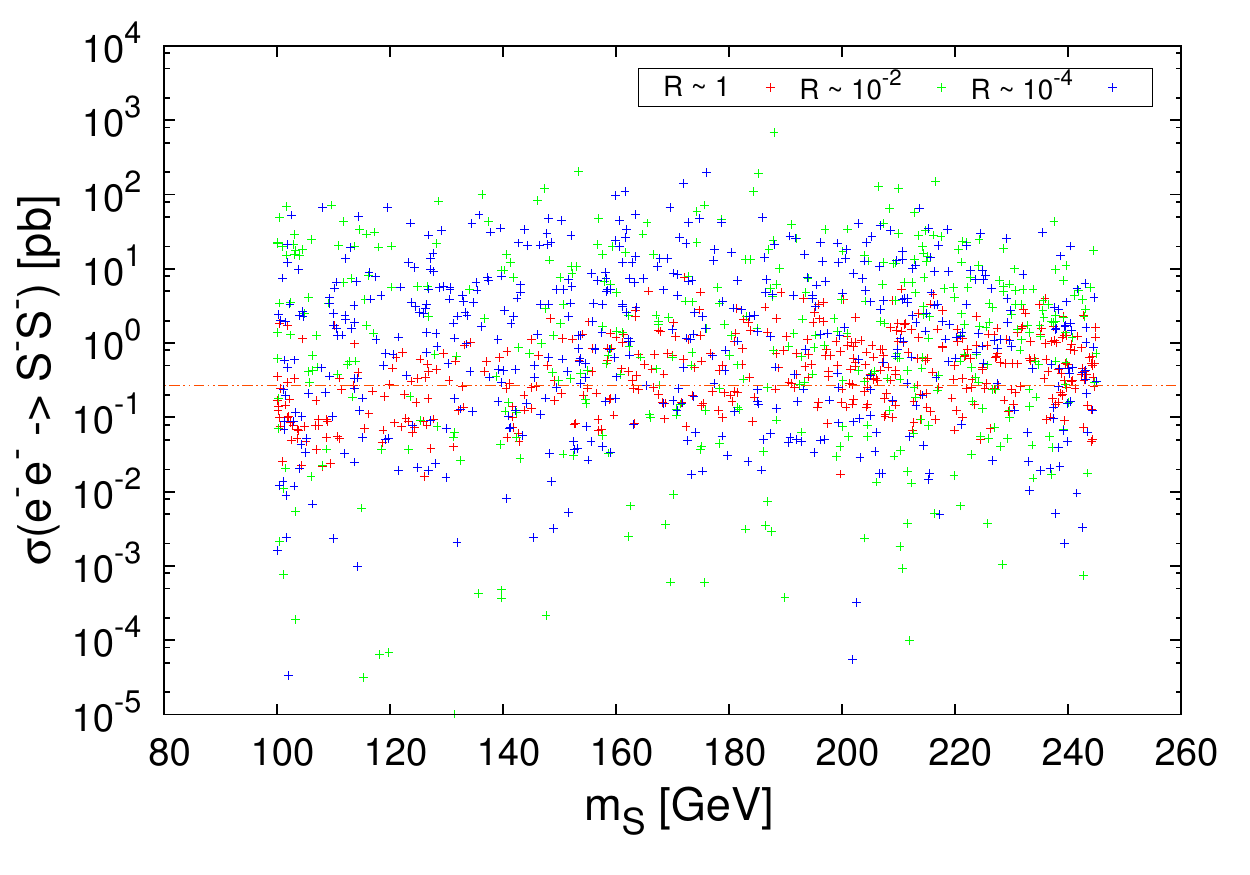}~\includegraphics[width=0.32\textwidth]{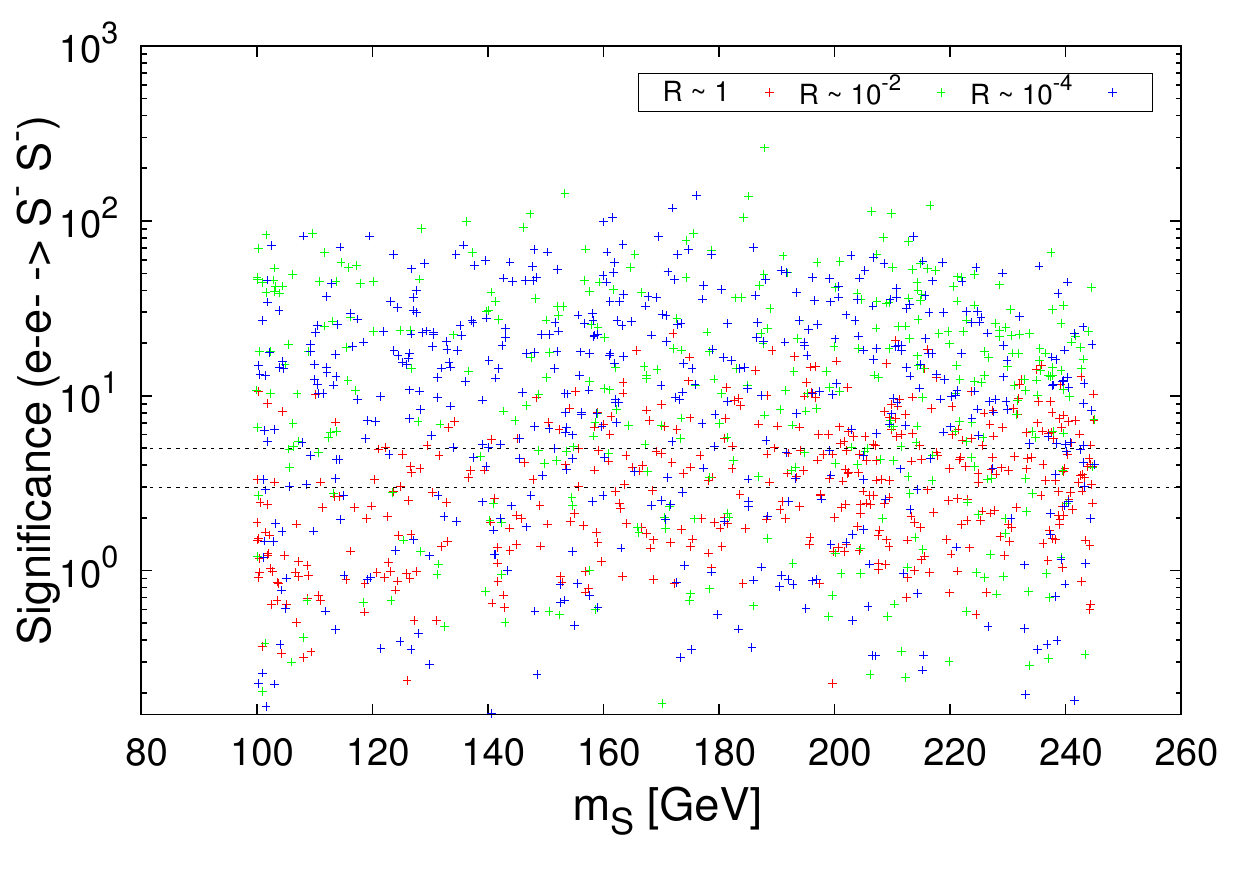} 
\par\end{centering}
\caption{The cross section values (top) and the corresponding significance values (middle) for production via electron-positron 
collision and at bottom for production via electron-electron collision at luminosity $100~pb^{-1}$ in function of $m_S$. The red 
lines represent the background value and the dashed one represents the Drell-Yann contribution in cross section, and the dashed 
lines represent $S=3,5$ in significance, respectively.}
\label{CS} 
\end{figure}}

\begin{table}[h]
\caption{Three benchmark points selected from the parameters space of the model.}
\begin{centering}
\begin{adjustbox}{max width=\textwidth}
\begin{tabular}{cccc}
\hline 
Point & $B_{1}\left(R_{1}\right)$ & $B_{2}\left(R_{2}\right)$ & $B_{3}\left(R_{3}\right)$\tabularnewline
\hline 
$g_{1e}$ & $(7.506+i0.014)\times10^{-1}$ & $(1.8284+i0.103)$ & $(-0.103+i0.201)$ \tabularnewline
\hline 
$g_{2e}$ & $(-0.26819-i1.5758)\times10^{-4}$ & $(1.543+i3.004)\times10^{-4}$ & $(0.654-i2.616)\times10^{-2}$\tabularnewline
\hline 
$g_{3e}$ & $(-1.360-i0.707)$ & $(0.313-i0.549)$ & $(-0.869-i0.878)$\tabularnewline
\hline 
$m_{S}$(GeV) & $196.75$ & $242.81$ & $104.47$\tabularnewline
\hline 
$m_{N_{1}}(\text{GeV})$ & $25.788$ & $43.764$ & $38.306$\tabularnewline
\hline 
$m_{N_{2}}(\text{GeV})$ & $28.885$ & $58.182$ & $56.481$\tabularnewline
\hline 
$m_{N_{3}}(\text{GeV})$ & $36.274$ & $67.511$ & $72.440$\tabularnewline
\hline 
\end{tabular}\end{adjustbox} 
\par\end{centering}
\label{tab-point} 
\end{table}

\vspace{2cm}
\hspace{-0.6cm} which occure via the interactions in (\ref{LL}), and give the following processes:

\begin{eqnarray}
e^{-}e^{+}&\rightarrow &\gamma+E_{miss}, \nonumber\\
e^{-}e^{+} &\rightarrow & S^{+}S^{-}\rightarrow\ell_{\alpha}^{+}\ell_{\beta}^{-}+E_{miss},\nonumber\\
e^{-}e^{-} &\rightarrow & S^{-}S^{-}\rightarrow\ell_{\alpha}^{-}\ell_{\beta}^{-}+E_{miss}\nonumber\\
e^{-}e^{+}&\rightarrow &\gamma+S^{+}S^{-}\rightarrow\gamma+\ell_{\alpha}^{+}\ell_{\beta}^{-}+E_{miss},\label{signal}
\end{eqnarray} 
Three processes for electron-positron collision are analyzed; photon(s) with a pair of DM in the final state where the background 
contributing to the signal give left-handed neutrinos with photon, and a pair production of charged scalars $S^{+} S^{-}$ with or 
without a photon in the final state, where each charged scalar decays into a RH neutrino and a charged lepton. The corresponding 
background comes from the process $e^{+}e^{-}\rightarrow W^{+}W^{-}$ where $W$ decays into a light neutrino and a charged lepton. 
Another potential signature come from electon-electron collison and give same sign pair of charged scalars. A first qualitative 
analysis is carried out on the four processes in (\ref{signal}), on three sets of benchmark points according to different values 
of the ratio $R$ for a center of mass energy of $\sqrt{s}=500~\text{GeV}$ and with a luminosity $L=100\,pb^{-1}$. The cross section 
values and the corresponding significance are shown in Fig.~\ref{CS} versus the charged scalar mass without applying any cut\footnote{Except the cut $E_{\gamma}>$ 8 GeV and $|cos~\theta_{\gamma}|<0.998$ on channels with photon in finale state.}. One remarks that the cross 
section values of the processes (\ref{signal}) in Fig.~\ref{CS} varies over seven orders of magnitudes as its sensitively depends on 
our choice of the parameters space. The production cross section via electron-electron collision is large compared to the background, 
so even for low luminosity the significance is huge. Hence, same sign pair of charged scalars process is a clean and RH neutrinos can 
be directly probed at ILC. Detectability of production via electron-positron collision is developed in more detail in the next chapter.

\setlength\textfloatsep{8pt plus 2pt minus 2pt}
\setlength\intextsep{8pt plus 2pt minus 2pt}
\begin{figure}[h]
\begin{centering}
\includegraphics[width=0.31\textwidth,height=0.145\textheight]{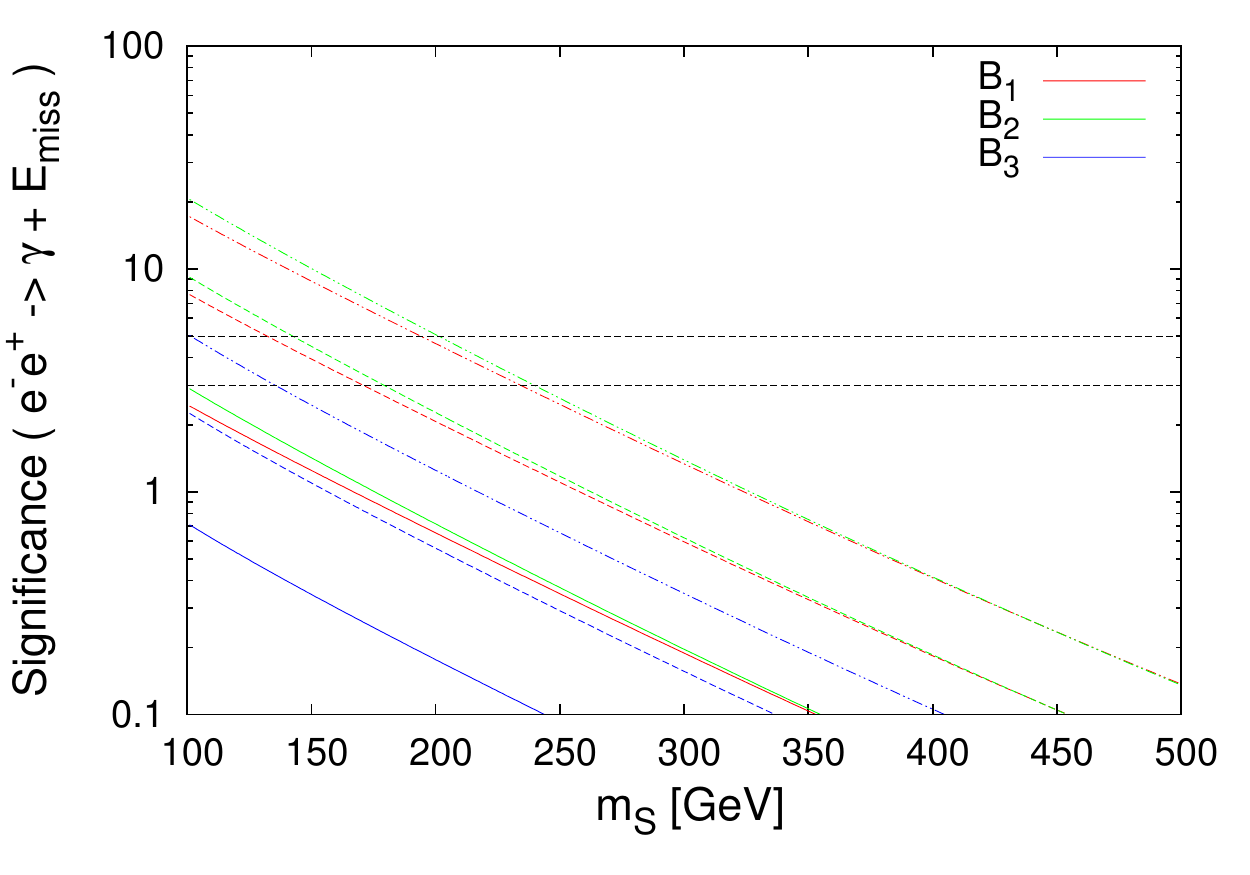}~\includegraphics[width=0.31\textwidth,height=0.145\textheight]{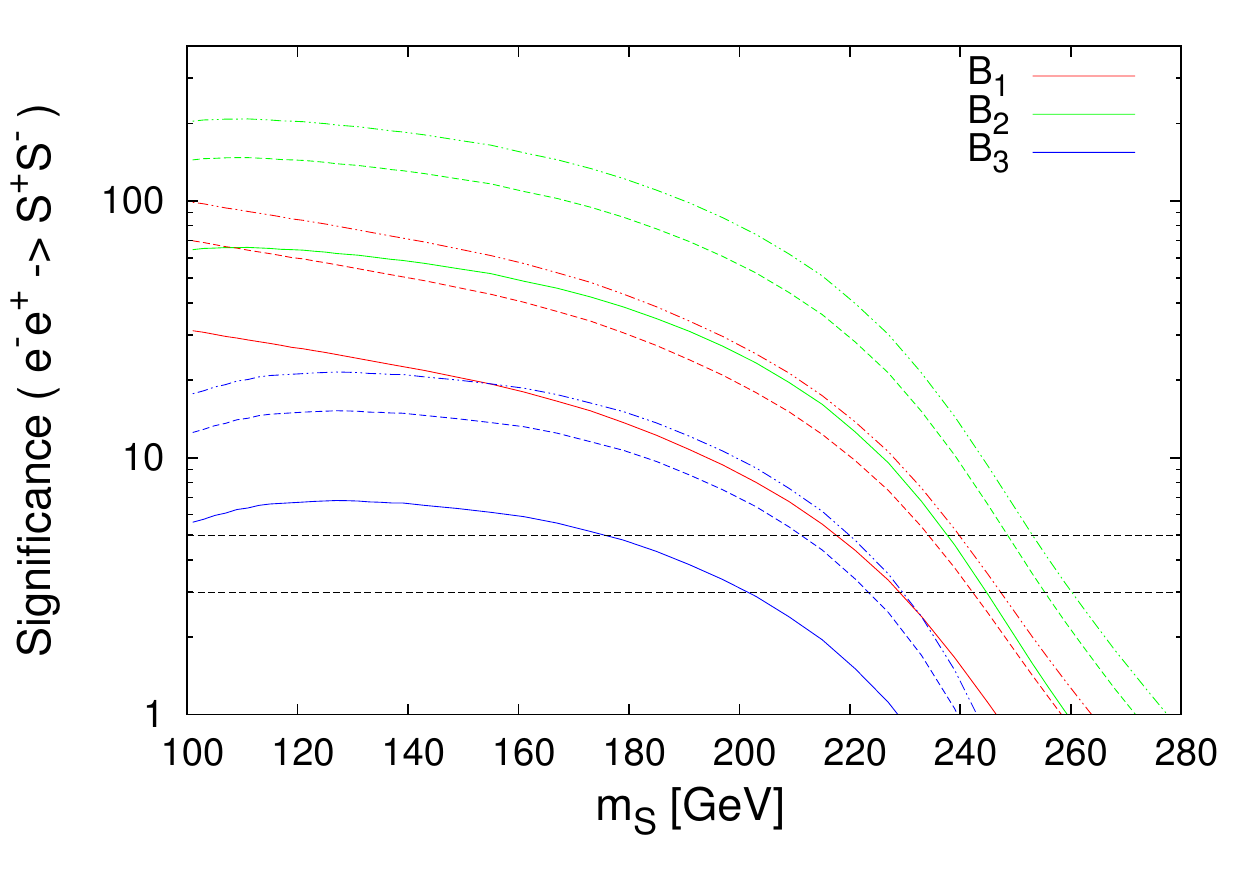}~\includegraphics[width=0.31\textwidth,height=0.145\textheight]{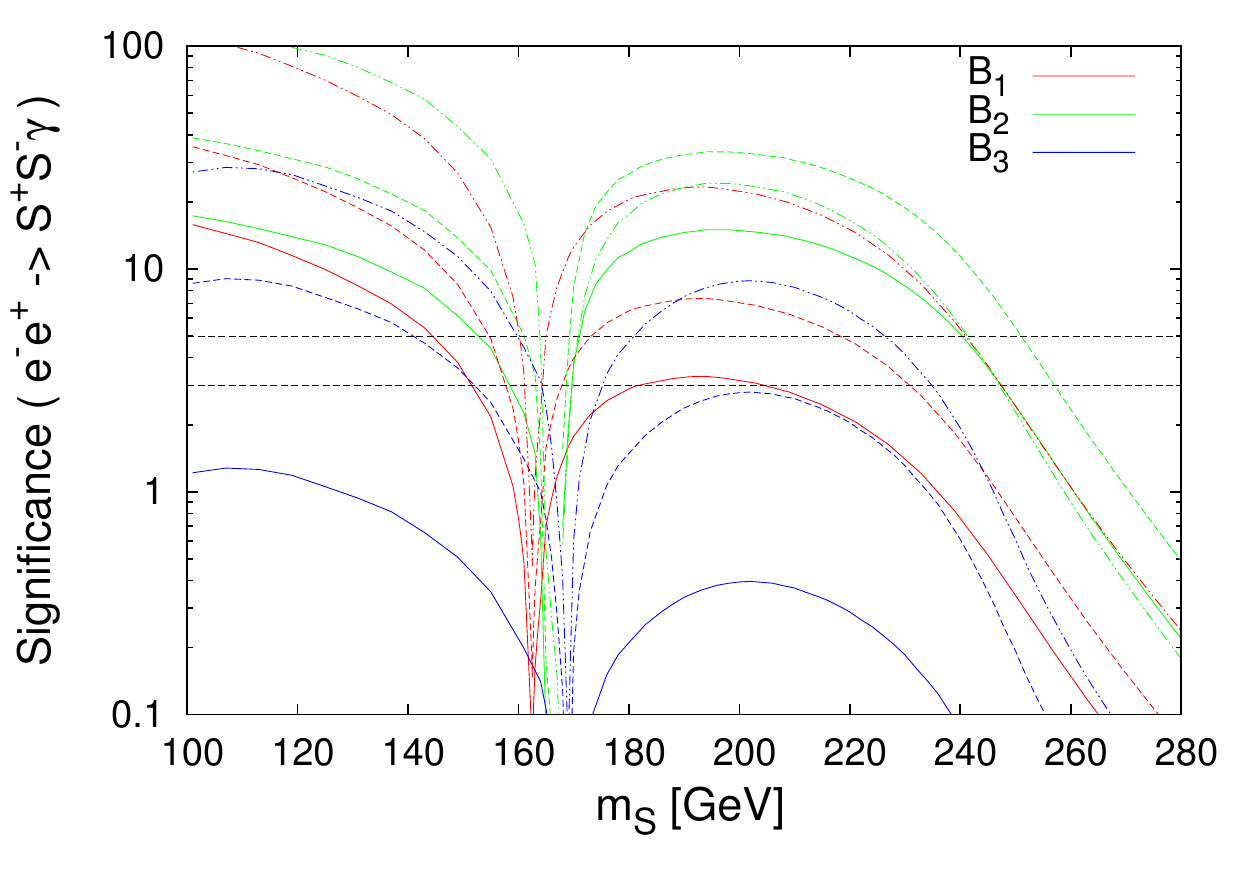} 
\par\end{centering}
\caption{The signal significance for the processes $e^{-}e^{+}\rightarrow\gamma+E_{miss}$ (left), $e^{-}e^{+}\rightarrow S^{-}S^{+}$ (middel) and $e^{-}e^{+}\rightarrow S^{-}S^{+}+\gamma$ (right) as a function of
$m_{S}$ for the values of $g_{i\alpha}$ given Table.~\ref{tab-point} at integrated luminosity of 1/10/10 (solid), 5/50/100 (dashed) 
and 10/100/500 (dash-dotted) $fb^{-1}$ respectively. The horizontal dashed lines correspond to a 3 and 5 sigma significance. For the 
values $m_S> 250~\text{GeV}$, the charged scalars is off-shell.}
\label{s-eES-A}
\end{figure}

\section{Benchmark Analysis}

Let's now consider three benchmarks points, one of each ratio $R_1\approx 1,~R_2\approx 10^{-2}$ and $R_3\approx 10^{-4}$, 
with nearby heavy neutrinos masses relatively. As can be seen on the table.\ref{tab-point}, our freedom of the model parameters 
space are substantially limited by the choice of the ratios $R_i$.The distributions for different kinematic variables are 
generated for signal and background Using CalcHEP~\cite{CalcHEP} for the processes $e^{-}e^{+}\rightarrow \gamma+E_{miss}$, 
$e^{-}e^{+}\rightarrow S^{-}S^{+}$, and $e^{-}e^{+}\rightarrow S^{-}S^{+}+\gamma$ at $500~\text{GeV}$. We extract the optimal 
kinematical cuts for each process and this can be achieved as the following,

\begin{center}
$\text{Final state}~\gamma + E_{miss}:~8~\text{GeV}<E_{\gamma}<300~\text{GeV},~|cos~\theta_{\gamma}|<0.998$\\
 $~\text{and}~E_{miss}>300~\text{GeV}.$
\end{center}

\begin{center}
$\text{Final state}~S^{+}S^{-}:~M_{\ell^{+},\ell^{-}}<300~\text{GeV},~150~\text{GeV}<E_{miss}<420~\text{GeV}$,\\
 $30~\text{GeV}<E^{\ell}<180~\text{GeV}~\text{and}~p_t^{\ell}<170~\text{GeV}.$
\end{center}

\begin{center}
$\text{Final state}~S^{+}S^{-} \gamma:~M_{\ell^{+},\ell^{-}}<300~\text{GeV},~150~\text{GeV}<E_{miss}<400~\text{GeV}$,\\
$~30~\text{GeV}<E^{\ell}<170~\text{GeV},~p_t^{\ell}<170~\text{GeV},|cos\left(\theta_{\gamma}\right)|<0.5$,\\ 
$8~\text{GeV}<E^{\gamma}<120~\text{GeV}~\text{and}~p_{t}^{\gamma}<110~\text{GeV}$.
\end{center}

By varying the charged scalar mass, the significance for these processes for the three considered benchmark points are shown in 
Fig.~\ref{s-eES-A} at integrated luminosity that brings us closer to 5 sigma significance for each channels. For the monophoton 
process, the signal

\begin{figure}[h]
\begin{centering}
\includegraphics[width=0.4\textwidth,height=0.18\textheight]{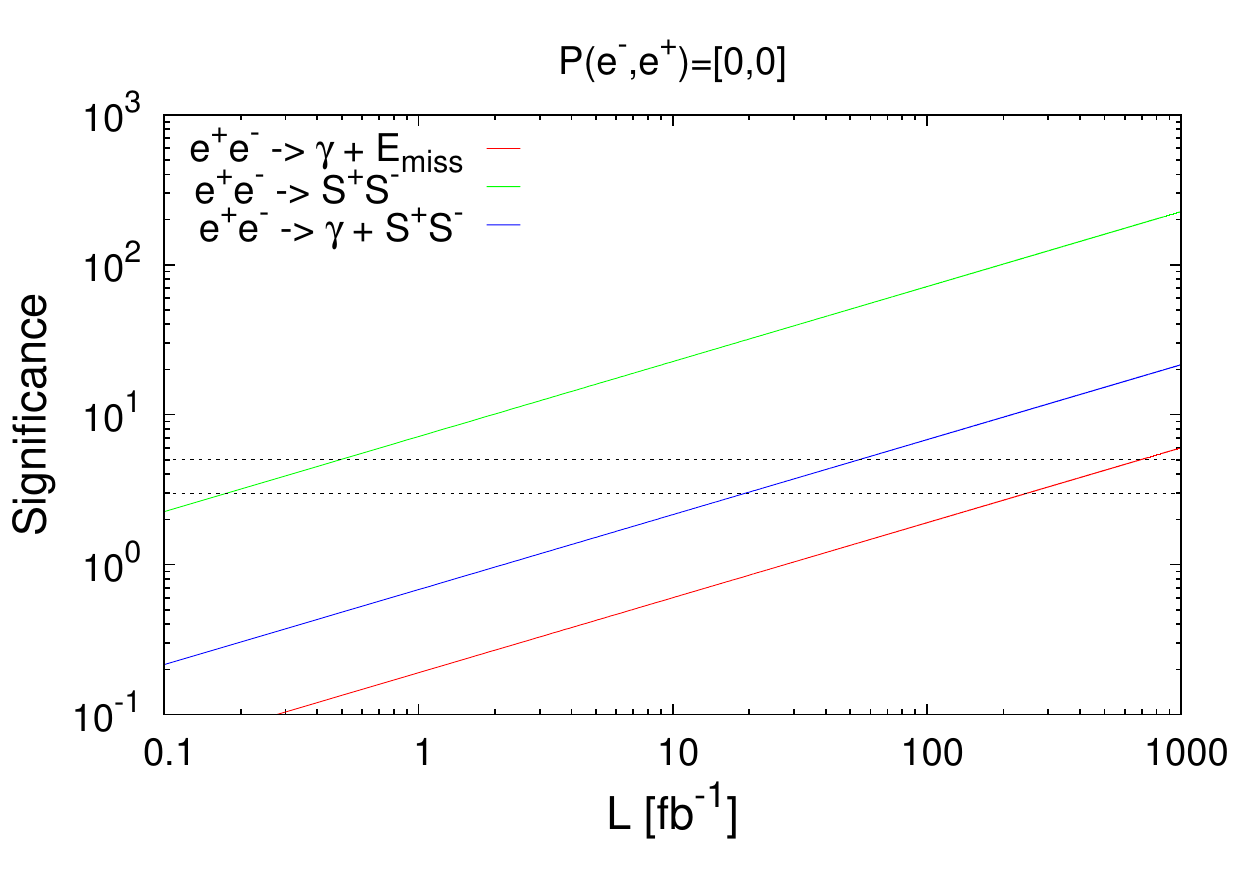}~\includegraphics[width=0.4\textwidth,height=0.18\textheight]{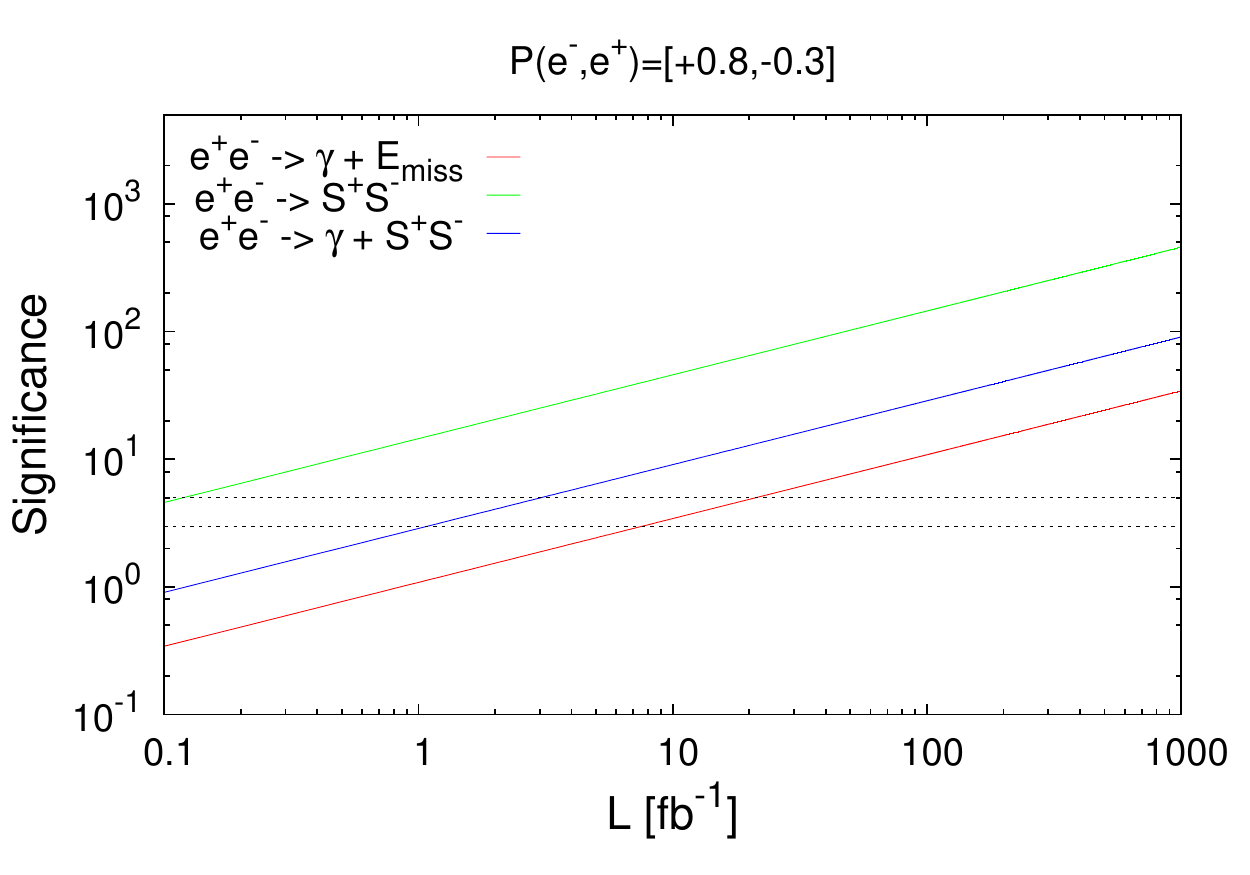} 
\par\end{centering}
\caption{The signal significance as function of luminosity for the three different signatures studied, at left without polarization and 
at right with polarization of $P(e^{-},e^{+})$=$[+0.8,-0.3]$. The two horizontal dashed lines correspond to to a 3 and 5 sigma 
significance, respectively.}
\label{sig-lum} 
\end{figure}

\hspace{-0.6cm} significance becomes detectable at the ILC for an integrated luminosity of a few hundred $\text{fb}^{-1}$. A 
luminosity of a few ten $\text{fb}^{-1}$ is necessary to probe the production of a pair charged scalars with a photon while the 
$S^{-}S^{+}$ channel could be visible easily at very low luminosity, around $0.5~\text{fb}^{-1}$. The charged scalar mass must be 
lighter than about 220 GeV in all these channels

There is an additional feature allowing the improvement of detection and which is available on the ILC, the possibility to have
highly polarized electron/positron beams. a longitudinal polarization of $80\%$ for the electron beam and $30\%$ for the positron 
beam are planned by the ILC. We re-analyze the processes discussed earlier with all the possible polarizations combinations in 
order to improve the signal-background ratio and we found that for polarized beams as $P\left(e^{-},e^{+}\right)=\left[+0.8,-0.3\right]$ 
while applying the same cuts used previously, the number of background events gets reduced by $86\%$ and the signal increased by $130\%$. 
In Fig.~\ref{sig-lum}, we present the significance for $P(e^{-},e^{+})$=$[0,0]$ and $P(e^{-},e^{+})$=$[+0.8,-0.3]$, for the benchmark 
point $B_3$ as a function of luminosity. one can note that the signal over background gets improved and the required integrated luminosity 
is dictated by a factor about ten for each processes studied.

\section{Conclusion}

This paper investigates the behavior of some type of interaction present in a class of models that extends the standard model by majorana 
right-handed neutrinos and sclaire charged, to explain violations of leptonic flavor on one side, and to give a serious condidat to the dark 
matter of another. After imposing several constraints on the free parameters of models and carrying out an in-depth analysis on the 
detectability of the major processes resulting from the electron-positron collision in the conditions of the future lepton collider : ILC, we 
show that this type of interaction is likely to be probed, for different luminosity according to the final state, but which are all largely 
within the scope of the ILC capacity. It is also shown that the polarization of the electon / positron beams present in this collider can lead 
to a positive result very quickly.

\begin{center}
*~~*~~*
\end{center}

MC wants to thank the organizers for the finanicial support. AA is supported by the Algerian Ministry of Higher Education and Scientific Research under the CNEPRU Project No B00L02UN180120140040.

\end{document}